\documentclass[12pt,aps,amsmath,latexsym,amsfonts]{JHEP3}

\usepackage{epsfig}
\usepackage{amsfonts,amssymb,amsmath}

\newcommand{\f}{\frac}

\newcommand{\lt}{\left}
\newcommand{\rt}{\right}

\newcommand{\al}{\alpha}
\newcommand{\dl}{\delta}
\newcommand{\ka}{\kappa}
\newcommand{\la}{\lambda}

\newcommand{\df}{\delta\phi}
\newcommand{\D}{[\mathcal{D}]}
\newcommand{\htt}{h^{{\tt TT}}_{\mu\nu}}

\newcommand{\be}{\begin{equation}}
\newcommand{\ee}{\end{equation}}
\newcommand{\bea}{\begin{eqnarray}}
\newcommand{\eea}{\end{eqnarray}}

\title{On black holes in heterotic braneworlds}
\author{Ruth Gregory, Armalivia Leksono and Bina Mistry \\
{\it Centre for Particle Theory, Department of Mathematical Sciences,}\\
{\it Durham University, South Road, Durham, DH1 3LE, United Kingdom}}

\abstract{
We explore the problem of braneworld black holes in the
heterotic braneworld scenario of Lukas, Ovrut, Stelle and Waldram (LOSW).
We show that black string solutions are unstable, and demonstrate
some unusual asymptotics in the linearized metric. We also solve
the fully coupled brane and bulk Einstein equations, finding an exact,
though singular, solution which corresponds to a brane black hole 
in which the branes spike apart at the Schwarzschild radius.
}

\preprint{DCPT-09/03}

\begin{document}

\section{Introduction}

Large extra dimensions and braneworlds have been an active topic of 
interest over the past decade, with many interesting implications
in phenomenology, cosmology, and gravity (see \cite{reviews} for 
reviews into these various aspects of LED's). While many
concrete predictions have been based on the explicit models of
Arkani-Hamed et.\ al.\ \cite{ADD}, or Randall Sundrum (RS) \cite{RS};
scenarios set in a string theory context such as KKLT \cite{KKLT}
or the heterotic braneworld \cite{LOSW} have also generated new
and interesting ideas in early universe cosmology. The heterotic
braneworld in particular has given rise to the ekpyrotic \cite{EKP},
and cyclic universe picture \cite{CYCU}, which has been the subject
of some controversy \cite{pyro}.

The heterotic braneworld is an interesting alternative to
type II string theory based models, and makes active use of the 
eleventh dimension to provide the ``large" extra dimension for our
braneworld. It is based on the Horava-Witten M-theory compactification
\cite{HW}, in which there is a hierarchical compactification to 4D
with the eleventh dimension larger than the remaining 6
spatial dimensions which are compactified on a Calabi-Yau (CY) manifold
\cite{HCOM}. The set-up then mimics the RS model, in which the 11th 
dimension plays the role of the distance normal to the brane.
The curvature of the CY sources wrapped M5 branes, which in turn
warp the ``fifth'' dimension in an analogous fashion to the RS model.

Despite the apparent similarities between the heterotic model and
the RS model, the presence of the scalar field in the gravitational
sector has a huge impact on the strong gravitational properties of
the braneworlds. In RS, cosmological braneworlds are precisely 
determinable \cite{RSCOS}, as the field equations are completely
integrable \cite{BCG}. In heterotic M-cosmology however, the
presence of the bulk scalar means that explicit analytic solutions 
can only be found by assuming an ansatz for the metric
\cite{MTHCOS}, and there is no ``Birkhoff'' theorem for the 
bulk\footnote{The ``modified Birkhoff theorem'' alluded to in
\cite{LMT} in fact makes a rather restrictive metric ansatz, and
cannot be taken as a general statement on the bulk spacetime. See
\cite{CCLiou}, \cite{CG} for general analytic analyses of more complex
situations.}.

Black holes are the other main test case for strong gravitational solutions,
and for the RS model, have proved to be rather problematic (see
\cite{BHrev} for a review). While the Schwarzschild solution on the brane
extends to a black string in the bulk \cite{CHR}, this string is
unstable \cite{BSINS}, and the exact solution is not known. Furthermore,
the parallels between the RS model and the gauge/gravity correspondence
of string theory \cite{RSADS} have led to the idea that a classical bulk 
solution will correspond to a quantum corrected black hole \cite{hol},
although the evidence so far is equivocal \cite{debate}. 
For the heterotic braneworld though, we do not expect a holographic 
analogy, indeed, it is unclear what
sort of black hole solution we can expect. Of course the Schwarzschild 
solution should provide a black string metric -- but is this a sensible
solution?  Here, we explore the properties of
heterotic brane black holes, determining what features the metric 
should have, and what the constraints on the system are. We show
the black string is unstable, then calculate the linearized black 
hole solution. We then comment on the full nonperturbative problem,
showing how, unlike RS, there is no approximate model for a mini
black hole metric, and explore a candidate brane plus bulk solution
with brane spherical symmetry.

\section{Overview of the heterotic braneworld and perturbation theory }

In this section, we describe the braneworld setup of Lukas, Ovrut,
Stelle and Waldram (LOSW) \cite{LOSW}, and give the background
solution of heterotic M-theory. We then derive the linearized Einstein and
scalar field equations.

\subsection{Heterotic M-theory}

We use the dimensionally reduced five-dimensional effective action 
consisting of 
a bulk scalar-tensor gravity, and two boundary branes:
\bea
S = &&\f{1}{2\ka_5^2}\int d^5x{\sqrt{-g}\lt[ -R +\f{1}{2}
(\partial\phi)^2- 6 \alpha^2e^{-2\phi}\rt]} \nonumber\\
&&+\f{6\alpha}{\ka_5^2}\lt[\int_{y=-y_0}d^4x{\sqrt{-g^-}e^{-\phi}}
-\int_{y=+y_0}d^4x{\sqrt{-g^+}e^{-\phi}}\rt], \label{eq:action}
\eea
where $R$ is the five-dimensional Ricci scalar, $g^{\pm}_{\mu\nu}$ is the
induced metric on each brane, $\kappa_5^2=8\pi G_5$ is the effective
five dimensional Newton's constant, and $\al$ is an arbitrary coupling constant,
parametrizing the number of units of 4-form flux which thread the 
Calabi-Yau\footnote{Note, for convenience, we are using the conventions
of \cite{LMT} for $\alpha$ rather than \cite{LOSW}.}.
The boundary branes have equal and opposite tensions and are positioned parallel
to each other at $y=\pm y_0$ (where $y$ is the transverse direction to the
brane), and we impose a $\mathbb{Z}_2$ symmetry at the position of each 
brane. 

The resulting five-dimensional equations of motion following from the action
(\ref{eq:action}) are
\be\begin{split}
G_{ab} =& \f{1}{2}\phi,_{a}\phi,_{b} - \f{1}{4}g_{ab}\phi,^{c}\phi,_{c}
+ 3 g_{ab}\al^2e^{-2\phi}\\
&- 6 \al[\dl(y+y_0)-\dl(y-y_0)]g_{\mu\nu}\dl^{\mu}_a\dl^{\nu}_b
\f{e^{-\phi}}{\sqrt{g_{yy}}} \label{eq:einsteineqn} 
\end{split}\ee
\be
\square\phi = 12 \al^2e^{-2\phi} - 12
\al[\dl(y+y_0)-\dl(y-y_0)]\f{e^{-\phi}}{\sqrt{g_{yy}}}, \label{eq:scalfieldeqn}
\ee
where Greek indices run over the four braneworld dimensions,
and Latin indices run over all five spacetime dimensions.

The heterotic braneworld is given by the solution (writing $y=x^{11}$)
\bea
\phi &=& \phi(y) = \ln H = \ln \f{[ 1 + 6 \alpha y]}{[1+6\alpha y_0]} \\
ds^2 &=& a^2(y)\eta_{\mu\nu}dx^{\mu}dx^{\nu} -dy^2 
= H^{1/3} \eta_{\mu\nu}dx^{\mu}dx^{\nu} -dy^2 
\label{background}
\eea
Note that this solution is in Gaussian Normal (GN)
gauge, ($g_{yy} = 1, g_{y\mu}=0$), 
a different gauge from the one originally written
down in \cite{LOSW}. This is primarily for calculational convenience,
and a general interacting brane system will 
need two coordinate patches, one for
each brane \cite{CGR}, in order to correctly encode the
physical degrees of freedom.
We have normalized the warp factor $a(y)$ so that $a=1$ on the positive
tension brane.

\subsection{Perturbations of the braneworld}

Following the usual braneworld procedure (see e.g.\ \cite{GT}), 
we write the perturbation
of the metric as $g_{ab} \to g_{ab}+h_{ab}$, and choose to remain
in the GN gauge. In addition, we would like to keep the coordinate
positions of the branes fixed at $\pm y_0$ for calculational
simplicity, which means that our coordinate system is no longer global,
and an explicit scalar degree of freedom is introduced
into the metric perturbation. The physical system consists of the bulk
and the two branes, and the physical degrees of freedom are therefore
bulk fluctuations and (potentially) a degree of freedom corresponding to
the fluctuation in position of each brane.
In the GN gauge, we encode this by explicitly performing a gauge transformation 
\be
y \to y+ f(x^\mu)\;, \qquad
x^\mu \to x^\mu + \frac{a^4}{ 4\alpha } \ \eta^{\mu\nu} f_{,\nu} 
\ee
which maintains the GN gauge, but shifts the brane to $y_0 + f$.
Since we have two branes, we can potentially have two different such
gauge transformations, and we require two gauge patches, one for each 
brane \cite{CGR}. In the overlap, we simply perform the relevant shift
of the $y$-coordinate to match the two patches.
Under such a change of coordinates, the metric and scalar field
change according to their Lie derivatives along the coordinate transformation:
\bea
\delta h_{\mu\nu} &=& \frac{a^6}{2\alpha} f_{,\mu\nu} + \frac{2\alpha}{a^4}
f \eta_{\mu\nu} \\
\delta \phi &=& f \phi' = 6f \frac{a'}{a}
\eea
This gives rise to an explicit scalar component in the perturbation.

After some algebra, the perturbation equations around the background
(\ref{background}) are found to be
\begin{eqnarray}
\f{1}{a^2}\lt[ a^2\lt(\f{h}{a^2}\rt)'\rt]' &=& - 2\phi'(\df)'
+8 \alpha^2 e^{-2\phi} \df-16\alpha e^{-\phi}\D\df 
- \f{2\kappa_5^2 T^\mp}{3a^2} \delta(y\pm y_0) \label{eq:hyy1} \\
\lt[\f{h_{\mu\la}{}^{,\la} -h,_{\mu}}{a^2}\rt]' &=& \phi'(\df)_{,\mu}
\label{eq:hmuy1}\\
\f{1}{a^2} \partial^2 h_{\mu\nu}
&-&\f{1}{a^2}\lt[a^4\lt(\f{h_{\mu\nu}}{a^2}\rt)'\rt]'
-  a^{-2} \lt( h_{,\mu\nu}-2h_{\la(\mu,\nu)}{}^{\la}\rt)
-aa'\lt[\f{h}{a^2}\rt]'\eta_{\mu\nu} \label{eq:hmunu1} \\
&=& -8 a^2 \alpha^2 e^{-2\phi} \eta_{\mu\nu} \df 
+ 4 a^2 \alpha e^{-\phi}\D \eta_{\mu\nu} \df
-2\kappa_5^2[T^\mp_{\mu\nu} - \f{T^\mp}{3} \eta_{\mu\nu} ] 
\delta(y\pm y_0) \nonumber\\
\f{\phi'}{2} \lt[\f{h}{a^2}\rt]' &=& +\f{\partial^2\df}{a^2} 
-(\df)'' -4\f{a'}{a}(\df)'
+24\alpha^2 e^{-2\phi} \df-12\alpha e^{-\phi}\D \df \label{eq:dilaton1}
\end{eqnarray}
where $\D=[\dl(y+y_0)-\dl(y-y_0)]$ represents the 
braneworlds, and have included a possible matter perturbation
$T_{_\mp\mu\nu}$ on each brane. (Note, spacetime indices $\mu,\nu...$
are raised and lowered by $\eta_{\mu\nu}$.)

The solutions to the homogeneous equations split naturally into
a zero mode sector, and a massive KK tower of tensor modes.
Inspecting (\ref{eq:hmunu1}) shows that the bulk dependence of
the zero mode sector must be either proportional to $a^2$, or
$a^2\int a^{-4} = a^4/2\alpha$. The first option of a profile
proportional to the warp factor gives 
a transverse-tracefree tensor mode, $\chi_{\mu\nu}$,  as the scalar
part of $h_{\mu\nu}$ is a pure 4D gauge mode. The other profile is only
consistent with a 4D scalar mode, and requires the 
presence of brane fluctuations in order to satisfy the boundary
conditions at each brane. This gives rise to a situation
in which we need two coordinate patches \cite{CGR}, in which
we have the GN gauge for each individual brane. The full zero mode
perturbation is given by
\begin{eqnarray} 
h_{_\pm\mu\nu} = a^2\chi_{\mu\nu} + \frac{f_{,\mu\nu}}{2\alpha}
\lt[ a^4 - \f{a^6}{2a_\pm^2}\rt] - \f{\alpha}{a_\pm^2a^4} f \eta_{\mu\nu} 
\label{TTFzero}\\
\delta\phi = -\f{3\alpha}{a_\pm^2a^6} f
\label{phizero}
\end{eqnarray}
which is interpreted as the massless graviton ($\chi$), and the 
radion ($f$). The $\pm$ subscript denotes the coordinate 
patches relevant to each brane. The first coordinate patch includes 
the brane located at $y=+y_0$ and the second includes the brane 
located at $y=-y_0$. Each coordinate patch is Gaussian Normal with 
respect to the brane it includes (but it does not have to be 
GN with respect to the other brane).  The transformation
on the overlap is read off from (\ref{TTFzero}) as:
\be
y \to y + \f{f}{2a_+^2} - \f{f}{2a_-^2}.
\ee
This represents fluctuations in the interbrane distance, and is 
a `breathing mode' for the fifth dimension.
Note that there is no additional scalar mode coming from the $\phi$-field,
as was originally discussed in \cite{BBDR}.
\TABLE{
\label{tab:KKmass}
\caption{A table of the first three KK mass eigenvalues for differing
values of $\alpha$ and $y_0$. $y_0$ is chosen to be either half, or very
nearly its maximal value $1/6\alpha$. Note that as $\alpha \to 0$, the lowest
eigenvalue, $m_0 \propto \alpha$, as can be seen from the asymptotic form
of the Bessel functions in (\ref{KKmasses}). 
By inspection, we see this is well approximated by $m_0\simeq 10\alpha$.}
\begin{tabular}{|l|l|l|l|} \hline
$\alpha = 1, y_0 = 1/12$ & 18.97 & 37.53 & 56.16  \\ \hline
$\alpha = 1, y_0 = 0.99/6$ & 10.01 & 19 & 27.92  \\ \hline
$\alpha = 1/2, y_0 = 1/6$ & 9.58 & 18.95 & 28.37  \\ \hline
$\alpha = 1/2, y_0 = 0.99/3$ & 5.02 & 9.5 & 13.96  \\ \hline
$\alpha = 0.1, y_0 = 9.9/6$ & 1.005 & 1.9 & 2.792  \\ \hline
$\alpha = 0.01, y_0 = 99/6$ & 0.1005 & 0.19 & 0.2792  \\ \hline
\end{tabular}
}

For the massive KK tower, $\partial^2 h_{\mu\nu} = - m^2 h_{\mu\nu}$,
and $\partial^2 \delta \phi = -m^2 \delta\phi$, and the scalar and tensor 
modes can be treated separately. 
Combining (\ref{eq:hyy1},\ref{eq:hmuy1},\ref{eq:dilaton1}) 
in the bulk leads to a third order equation for 
the `scalar' perturbation:
\be
\f{1}{a^2} \partial^2 \left ( a^6 \delta \phi \right )' - a^6
\left [ \f{(a^6 \delta \phi)''}{a^6} \right ] ' 
-20 \left ( \f{a'}{a} \right )^2 \left (a^6 \delta\phi\right )'
=0
\ee
Clearly, $\delta \phi \propto a^6$ corresponds to the zero mode already
discussed, hence there are only two other possible solutions
\be
(a^6 \delta\phi)' \propto J_{4/5} (mz) \;\; , \;\; J_{-4/5} (mz)
\ee
where 
\be
z = \int \f{dy}{a} = \f{a^5}{5\alpha}
\label{zofy}
\ee
is a conformal transverse
coordinate. However, (\ref{eq:hyy1}) and (\ref{eq:dilaton1}) imply that
at each brane
\bea
\left ( \f{h}{a^2} \right ) ' &=& -8\alpha e^{-\phi} \delta \phi \\
(\delta\phi)' + 6\alpha e^{-\phi} \delta\phi &=&0
\eea
While we can balance the coefficients of the fractional Bessel functions
to satisfy this second equation, the first is not consistent
with (\ref{eq:dilaton1}) in a neighborhood of the boundary for $m^2 \neq 0$.

We are thus left with the massive spin 2 modes
$\htt = u_m(y) \chi^{(m)}_{\mu\nu}$, with
\bea
0 &=& - u_m \partial^2 \chi^{(m)}_{\mu\nu} + \chi^{(m)}_{\mu\nu}
\left [ a^4 \left ( \f{u_m}{a^2} \right )' \right ]' \nonumber \\
&=& a^3 \chi^{(m)}_{\mu\nu} \left [ \f{d^2}{dz^2} \f{u_m}{a^3}
+ \f{1}{z} \f{d}{dz} \f{u_m}{a^3} + m^2 \f{u_m}{a^3} - \f{1}{25z^2}
\f{u_m}{a^3} \right ]
\eea
thus $u_m \propto a^3 J_{\pm 4/5}\left(m\,z(y)\right)$, and we have
\begin{equation}
u_m  = a^3 \sqrt{\f{m}{5\alpha}}
\f{\lt[J_{4/5}(m z_+)J_{1/5}(mz)+J_{-4/5}(m z_+)J_{-1/5}(mz)\rt]}
{\sqrt{J^2_{-4/5}(m z_+)+J^2_{4/5}(m z_+)}}. \label{eq:masseigenval}
\end{equation}
Note this normalization corresponds to a continuum normalization, and
as such is only strictly valid in the limit $\alpha \to 0$, $y_0 \to \infty$,
with $\alpha y_0$ fixed. However, for sufficiently small $\alpha$ it gives 
a good working approximation for the computation of the propagator, and is
far less unwieldy than the exact expression involving hypergeometric
functions.
This eigenfunction explicitly satisfies the boundary conditions at the
`$+$' brane, however, in order for the boundary conditions at the second
`$-$' brane to be satisfied, we must have (see also \cite{SK}):
\be
J_{4/5}(m z_-)J_{-4/5}(mz_+)=J_{-4/5}(m z_-)J_{4/5}(mz_+)
\label{KKmasses}
\ee
This restricts the values of $m$ allowed in a fashion dependent on
$\alpha$ and the background interbrane distance $2y_0$. 
Some sample mass eigenvalues are shown in Table \ref{tab:KKmass}.
\FIGURE{
\label{fig:efns}
\includegraphics[height=4cm]{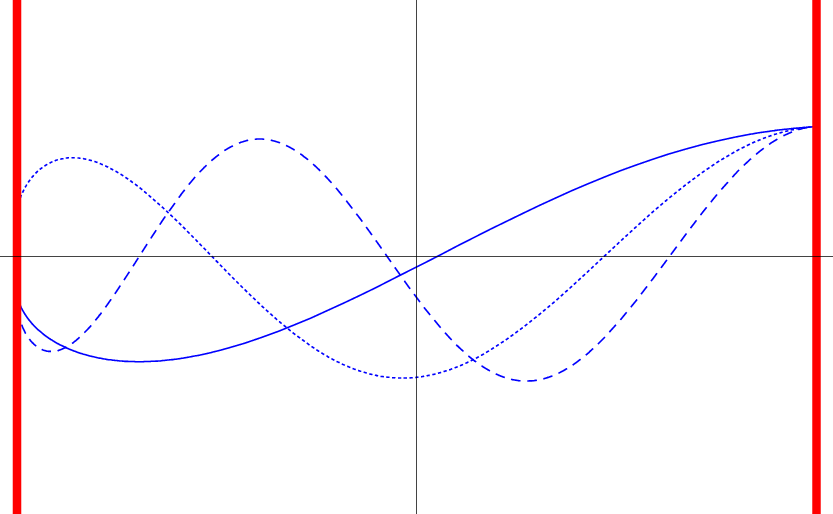}\hskip 5truemm
\nobreak\includegraphics[height=4cm]{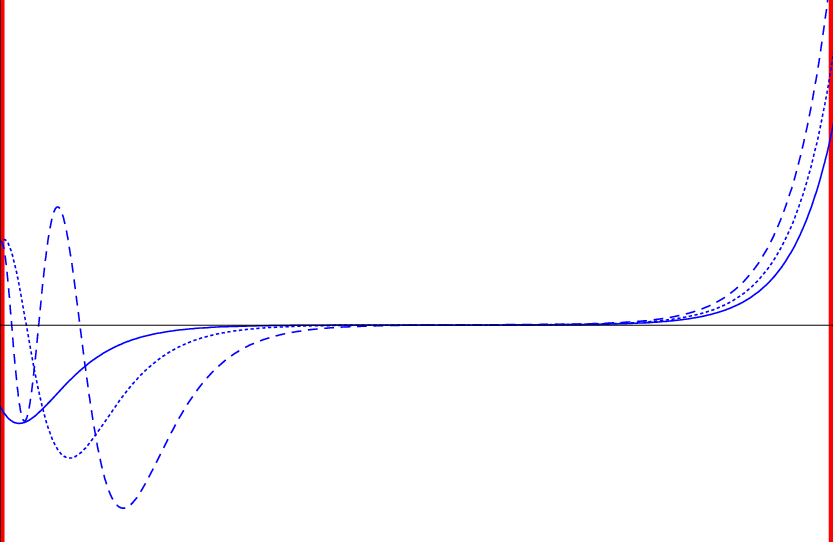}
\caption{A plot of the first three KK modes for $\alpha = 0.01$,
we have taken the branes (shown in red) to be at $\pm y_0 = \pm 33/2$.
The masses of the KK modes are given in the final line of 
Table \ref{tab:KKmass}. On the right, for comparison, are the first
three eigenfunctions for the RS model.}}

Like the RS scenario, the behaviour of the massive KK tower is 
determined by Bessel functions, although in this case they are
fractional Bessel functions.  Figure \ref{fig:efns} shows the first
three eigenfunctions, and for comparison the corresponding
RS profile. The heterotic eigenfunctions have a much more regular profile
along the $y$ direction, which is largely due to the power law,
rather than exponential, dependence of the warp factor on $y$. 

Drawing this information together, the Green's function for the spin 2
part of the perturbation on 
the heterotic braneworld is given in the continuum limit by:
\begin{equation}
G_R(x,x')= \f{4\alpha a^2(z) a^2(z')}{a^8_+} D_0(x-x') 
+\int_0^{\infty} dm \ u_m(z) u_m(z') D_m(x-x')
\end{equation}

This now allows us to compute the effect in the (positive tension)
brane of a source on the brane
\be
T^+_{ab} = \delta(y - y_0) T^+_{\mu\nu} (x^\mu) \delta^\mu_a \delta^\nu_b
\ee
The presence of a source on the brane will in general
require the introduction of a nonzero scalar perturbation, 
and once again we need two coordinate patches, one for each brane. 
This is a well known result from RS braneworlds, and is interpreted
as the brane bending in response to the matter source \cite{GT}. 

The general perturbation has the form:
\bea
h^+_{\mu\nu} &=& \htt + \f{a^6}{2\alpha} F^+_{,\mu\nu} + \f{2\alpha}{a^4} F^+
\eta_{\mu\nu} \\
\delta \phi^+ &=& \f{6\alpha}{a^6} F^+
\eea
where $F^+$ is the brane bending term in the coordinate patch of 
the `+' brane, and is given by:
\be
\partial^2 F^+ = \kappa_5^2 \f{T^+}{6}
\ee

Pulling all the information together, we see that the metric in the 
brane produced by matter on the brane is, up to a gauge transformation:
\bea
h^+_{\mu\nu} = &-& 8\alpha \kappa_5^2 \int d^4x'\; D_0(x-x') \left [
T^+_{\mu\nu} - \f{3}{8} T^+ \eta_{\mu\nu} \right ] \nonumber\\
&-& \kappa_5^2 \int d^4x' \int dm \; u^2_m(y_0) D_m(x-x') 
\left [ T^+_{\mu\nu} - \f{T^+}{3} \eta_{\mu\nu} \right]
\label{gravprop}
\eea
where 
\be
u_m^2(y_0) = \f{m}{5\alpha} \; \f{ \left[ J_{4/5}(\f{m}{5\alpha})
J_{1/5}(\f{m}{5\alpha}) + J_{-4/5}(\f{m}{5\alpha})
J_{-1/5}(\f{m}{5\alpha}) \right ]^2}
{\left [J_{4/5}^2(\f{m}{5\alpha}) + J_{-4/5}^2(\f{m}{5\alpha})\right ]}
\ee
and the dilaton is given by:
\be
\delta \phi^+ =  \alpha \kappa^2_5 \int d^4 x' \; D_0(x-x') \; T^+
\label{linscalar}
\ee
Thus the brane gravity is a Brans-Dicke theory with $\omega = 1/2$.

We now explore the possibilities for a heterotic brane black hole.
While we do not have a complete answer to this problem, there are
several approaches using both perturbation theory, as well as exact
solutions. We start by constructing the black string, and 
exploring its r\'egime of stability. Continuing the theme of linearized
theory, we compare this with the leading order solution for a point
particle on the brane. Then we turn to exact approaches, first discussing
the possibility of trajectories in known bulks before looking at the
full axisymmetric problem and presenting a possible (singular) solution.

\section{The Black String and Perturbation Theory}

A natural first step in looking for a brane black hole solution 
is to construct the black string:
\begin{equation}
ds^2 = a^2 \left [ \left ( 1 - 2\frac{G_NM}{r} \right ) dt^2
- \left ( 1 - 2\frac{G_NM}{r} \right )^{-1} dr^2 - r^2 d\Omega^2_{I\!I}
\right ] -dy^2 
\label{blackstring}
\end{equation}
However, based on our intuition of cylindrical types of horizon, we expect
this black string to be unstable \cite{GL}. The instability of the black 
string in KK theory occurs because there is an unstable massive tensor mode
of the 4D Schwarzschild metric, therefore, provided a the perturbations
of a spacetime allow a separation of the perturbation 
into an effective 4D tensor mode with an orthogonal mass eigenfunction, the
instability of the string persists to more complicated spacetimes.

It is not difficult to compute the perturbation equations around the
curved 4D background, and as in \cite{BSINS} we obtain for the 4D TTF
mode:
\be
a^{-2} \left ( \Box^{(4)} h_{\mu\nu} + 2 R^{(4)}_{\mu\lambda\nu\rho} 
h^{\lambda\rho} \right ) - h''_{\mu\nu} + 2 \left ( \frac{a''}{a} 
+ \left(\frac{a'}{a}\right)^2 \right ) h_{\mu\nu} = 0
\ee
where $a(y)$ takes the appropriate form for 
the heterotic background (\ref{background}).
As this is a tensor mode the scalar perturbations are not
excited, at least to linear order.
Thus, we can read off the unstable tensor mode as
\be
u_m(y) h^{(GL)}_{\mu\nu}(t,r)
\ee
where $h^{(GL)}_{\mu\nu}(t,r)$ is the unstable mode:
\be
h^{(GL)}_{\mu\nu} = e^{\Omega t} \left [ 
\begin{matrix} h_0 & h_1 & 0 & 0 \cr
h_1 & h_2 & 0 & 0\cr 0 & 0 & K & 0 \cr 0 & 0 & 0 & K\sin^2\theta \end{matrix}
\right ]
\ee
and $h_0,h_1,h_2$ and $K$ are all related via the TTF gauge conditions, and 
are given in \cite{GL}. The parameter $\Omega$ depends on the mass $m$ of the 
longitudinal tensor $s$-wave, and is found numerically \cite{GL}, however, 
for the 4D Schwarzschild metric it is well approximated by:
\be
\Omega(m) = \frac{m}{2} - m^2 G_N M \; .
\ee
Clearly there will be an unstable mode if the mass of the black string is 
(roughly) less than $1/2G_N m_0$, where $m_0$ is the minimum 
eigenvalue permitted for the massive tensor tower.
This minimum value depends on $y_0$ and $\alpha$, but for $y_0 \sim 1/6\alpha$,
i.e.\ if the branes are close to their maximal separation, the value is
well approximated by $m_0 \simeq 10\alpha$. Thus the onset of the
instability is given by
\be
G_N M \leq \frac{1}{20\alpha}
\ee

\FIGURE{
\label{fig:instability}
\includegraphics[width=7cm]{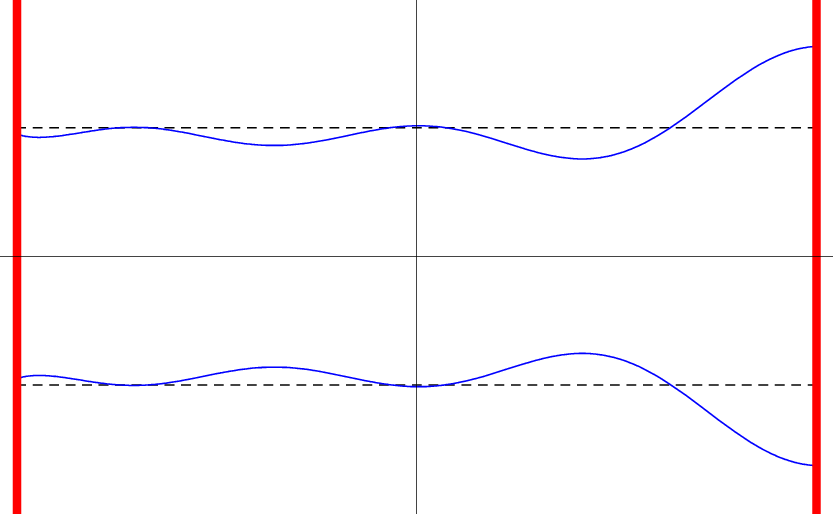}
\caption{The evolution of the black string instability
for the nearly marginal case of $G_NM=1$, $\alpha=0.01$, $y_0=99/6$. Here only
5 eigenfunctions are below the critical mass value, and the evolution of the
event horizon is exaggerated for clarity. The unperturbed event horizon
is shown as a dotted black line.}}
Once the instability has set in, the evolution is similar to the KK
string and is shown in figure \ref{fig:instability}. As with the RS
braneworld, the instability is focussed on the brane itself, however,
in this case the ripples in the event horizon across the bulk are more
uniform, mirroring the behaviour of the transverse eigenfunctions.
This does not give any reliable indication of the nonperturbative behaviour,
however it is reassuring that the instability does cluster near the brane, 
rather than having some strong bulk behaviour. One might therefore  expect
that the true black hole solution would be localized near the brane.

One can also use linearized theory to obtain a far-field approximation
of the black hole metric, by computing the 
linearized solution for a point source on the brane
\be
T_{\mu\nu} = M \delta ({\bf r}) \delta (y - y_0) \delta_\mu^0 \delta_\nu^0
\ee
For the scalar, (\ref{linscalar}) immediately gives
\be
\phi(r) = \f{2\alpha G_5 M}{r}
\ee
on the brane. For the tensor,
using (\ref{gravprop}), and expanding the Bessel functions in $u_m(0)$
at small $m$ gives a Newtonian potential of
\be
V(r) = - \f{10 \alpha G_5 M}{r} \left ( 1 + \f{2^{7/5} \Gamma[\f{8}{5}]}
{(5\alpha r)^{8/5} 3 \Gamma[\f{4}{5}]^2} \right )
\ee
This is quite an unusual potential because of the presence of the 
fractional powers of $r$. Note that it was obtained using the continuum 
approximation and therefore is only really valid for very small values
of $\alpha$, and large values of $y_0$.

\section{General axisymmetric bulk and the black hole solution}

Turning from perturbation theory, a natural approach
towards constructing a brane black hole is to use 
a known bulk and explore possible brane trajectories. Normally,
one uses a spherically symmetric bulk, taking the brane trajectory
through the bulk at some nontrivial trajectory $\theta(r)$, thereby
giving rise to a spherically symmetric brane.
Typically, although these trajectories exist, they do not correspond
to `empty' branes, and energy momentum is required on the brane to source
the gravitational field. These trajectories are solutions
of the brane Tolman-Oppenheimer-Volkoff equations (in the context of
the RS model, see \cite{BTOV} for work on brane stars and TOV, and
\cite{Seahra} for brane and bulk solutions).

Unfortunately, with the bulk action given by (\ref{eq:action}), there
are no spherically symmetric black hole solutions which are asymptotically
flat or (a)dS \cite{Wilt}. Indeed, even if one asks for only
planar symmetric solutions, the system of Einstein equations has lost
its simplicity, and only limited analytic information can
be extracted \cite{CCLiou}. Spherically symmetric geometries
exist only for special values of $\alpha^2$, and have unusual 
asymptotics \cite{CHM}. In the case of the heterotic braneworld, 
the sign of the Liouville potential does not permit such a solution, as
can be readily seen by attempting to solve (\ref{eq:einsteineqn}).
It may seem strange that there is no black hole solution, since the geometry
has arisen from a compactification from 11 dimensions. However, the
wrapped 5 branes, which give rise to the Liouville potential, mean that the
dilaton cannot remain fixed in the bulk, and not only remove the possibility 
of an asymptotically flat solution, but imply an anisotropy in any bulk
solution.

It seems therefore that to find a solution, a genuinely axisymmetric
bulk metric is required. Using coordinate freedom, this metric can
be expressed (up to a 2D conformal gauge group) as \cite{CG}:
\be
ds^2 = e^{2\sigma(r,z)} dt^2 - e^{2\chi(r,z)-\sigma(r,z)} B(r,z)^{-1/2} 
(dr^2 + dz^2) - B(r,z) e^{-\sigma(r,z)} d\Omega^2_{I\!I}
\ee
This metric has the Einstein equations:
\bea
\Delta B &=& \left [ 2 e^{2\chi} - 6 \alpha^2 B e^{2\chi-2\phi-\sigma} 
\right] B^{-1/2} \label{Baxieqn}\\
\Delta \sigma + \nabla \sigma \cdot \frac{\nabla B}{B} &=&
-2 \alpha^2 e^{2\chi-2\phi-\sigma}  B^{-1/2}\\
\Delta \chi + \frac{3}{4} (\nabla \sigma)^2  + \f{1}{4} (\nabla\phi)^2
&=& - \frac{e^{2\chi}B^{-3/2}}{2}
- \frac{3\alpha^2}{2} e^{2\chi-2\phi-\sigma} B^{-1/2}\\
\frac{\partial^2_\pm B}{B} + \frac{3}{2} (\partial_\pm \sigma)^2 
+ \f{1}{2} (\partial_\pm\phi)^2
- 2 \partial_\pm \chi \frac{\partial_\pm B}{B} &=& 0
\eea
and for the dilaton:
\be
\Delta \phi + \nabla \phi \cdot \frac{\nabla B}{B} =
-12 \alpha^2 e^{2\chi-2\phi-\sigma}  B^{-1/2}
\ee
where $\Delta$ is the 2D Laplacian on $(r,z)$ space, with $\nabla$ as 
the 2D gradient, and $\partial_\pm = \partial_r \pm i \partial_z$.

This system is similar to the axisymmetric spacetimes explored in \cite{CG}, 
where the general axisymmetric Einstein equations were derived, then analyzed 
in detail for the case of either spherical symmetry, or 
a cosmological constant. In each case three
classes of analytic solution were found. It was noted however, that 
these were specialized solutions, derived assuming some (albeit minimal)
metric Ansatz, and did not represent the full 
range of possibilities for the spacetime.

In this heterotic case, with both spherical symmetry and the bulk scalar field,
the set of equations is more involved, and like the Einstein axisymmetric
problem, does not have a general solution generating method. Interestingly
however, a simple (and commonly used) Ansatz of separation of metric
variables gives just one family of solutions.
Setting
\bea
B &=& b_1(r) b_2(z) \\
\sigma &=& \sigma_0 + \sigma_1(r) +  \sigma_2(z) \\
\chi &=& \chi_0 + \chi_1(r) + \chi_2(z) \\
\phi &=& \phi_0 + \phi_1(r) + \phi_2(z)
\eea
and inspecting (\ref{Baxieqn}) suggests that $e^{2\chi} B^{-1/2}$ is a function
of $r$, and $e^{2\chi-2\phi-\sigma} B^{-1/2}$ is a function of $z$. Other
possibilities are that the roles of $r$ and $z$ are swapped (which would 
result in a rotation of the branes), or both are
a function of $r$ (or $z$). Since it is the first option which corresponds
to the LOSW vacuum, we will use this in order to obtain asymptotically flat
braneworld solutions.

Using the restrictions on $\chi$ corresponding to $e^{2\chi} B^{-1/2}$
being a function of $r$, and $e^{2\chi-2\phi-\sigma} B^{-1/2}$
a function of $z$, the equations
of motion give the following expressions:
\bea
B &=& f(z) / g'(r) \\
\sigma &=& \sigma_0 + a g(r) +  \f{\ln f(z)}{3} - \f{bc}{3} \zeta \\
\chi &=& \chi_0 + \frac{(2b+1)a}{2} g(r) - \frac{1}{4} \ln g'(r)
+ \frac{3}{4} \ln f(z) \\
\phi &=& \phi_0 + ab g(r) + 2\ln f(z) + c \zeta 
\eea
where $a$, $b$, and $c$ are arbitrary constants, and 
\be
\zeta = \int \f{dz}{f}\; .
\ee
The Einstein equations give a pair of NLDE's for $f$ and $g$:
\bea
\f{\ddot f}{f} &=& - \f{2{\dot f}^2}{3f^2} + \left ( \f{b}{3}-2\right) 
c\f{\dot f}{f^2} - \f{(b^2+3)c^2}{6f^2} = 6\alpha^2 f^{-10/3}
e^{-(2-b/3)c\zeta} \\
\left ( \frac{1}{g'} \right ) ''  &=& 
\f{g'}{2} \left ( \f{1}{g'} \right )^{\prime 2} + (2b+1) a g'
\left ( \f{1}{g'} \right )' - \f{(3+b^2)a^2}{2} g' 
= 2 e^{2\chi_0} \ e^{(2b+1)ag}
\eea
where a dot denotes $d/dz$ and a prime $d/dr$.
The $f$ equation has the solution
\be
f = \left [ \f{2\alpha}{c\beta} \sinh (c\beta\zeta) 
e^{(b-6)c\zeta/6} \right]^{3/2}
\ee
where $\beta^2 = \f{2}{3} ( 1 - \f{b}{2} - \f{b^2}{8})$.
The $g$ equation can be integrated by making a change of variable:
\be
\rho = \int e^{(2b+1)ag}
\ee
which gives
\be
g(r) = \frac{1}{2E} \ln \left [ \frac{\rho-2E}{\rho} \right]
 = \f{1}{2E} \ln V_s(\rho)
\ee
where $E^2 = a^2(1+b+5b^2/4)$, and $V_s$ is of course the standard
4D Schwarzschild potential. 

Pulling this information together, we see that the general bulk separable
solution is:
\bea
ds^2 &=& f^{\f{2}{3}} e^{\f{bc\zeta}{3}} \left [ 
V_s(\rho)^{\f{a}{E}} e^{-bc\zeta} dt^2 
- V_s(\rho)^{-\f{(1+b)a}{E}} [ d\rho^2 + \rho(\rho-2E)
d\Omega^2 ] - f^2 V_s(\rho)^{\f{ab}{E}} d\zeta^2 \right ] \nonumber \\
e^{2\phi} &=& V_s(\rho)^{\f{ab}{E}} f^4 e^{2c\zeta}
\label{genbulksoln}
\eea
The LOSW vacuum corresponds to $a=c = 0$. In this case $E=0$ 
and all the nontrivial $\rho$-dependence drops out leaving us with
\be
ds^2 = 2\alpha\zeta \eta_{\mu\nu} dx^\mu dx^\nu - (2\alpha\zeta)^4 d\zeta^2
\ee
setting $6\alpha y = (2\alpha\zeta)^3-1$ recovers the original GN form.

Taking $b=c=0$, but $a = E\neq0$ recovers the ``uniform black string'' 
solution. For $b\neq0$ however, $a$ is no longer equal to $E$, and
the metric and scalar react to the ``source",
leading to the metric
\be
ds^2 = a^2(y) \left [ \left ( 1-\f{2E}{\rho} \right )^{\f{a}{E}} dt^2
- \left ( 1-\f{2E}{\rho} \right )^{-\f{a(1+b)}{E}}  [ d\rho^2 + \rho(\rho-2E)
d\Omega^2 ] \right ] - \left ( 1-\f{2E}{\rho} \right )^{\f{ab}{E}} dy^2 
\label{brnbulksoln}
\ee
with the scalar given by:
\be
e^{2\phi} =  \left ( 1-\f{2E}{\rho} \right )^{\f{ab}{E}} a^{12}(y)
\label{brnbulksclr}
\ee
The $y$ co-ordinate is no longer a GN coordinate because of the variation
of $g_{yy}$ in $\rho$.

Turning to a braneworld solution, we introduce branes at $\zeta = \zeta_\pm$, 
and compute the extrinsic curvature in order to evaluate the boundary
conditions:
\bea
K_{tt} &=& -f^{2/3}e^{(b-6)c\zeta_\pm/3}e^{-\phi} 
\left ( \f{f_{,\zeta}}{3f} - \f{ac}{3} \right ) g_{tt}\\
K_{\rho\rho} &=& -f^{2/3}e^{(b-6)c\zeta_\pm/3}e^{-\phi} 
\left ( \f{f_{,\zeta}}{3f} + \f{ac}{6} \right ) g_{\rho\rho}\\
K_{\theta\theta} &=& -f^{2/3}e^{(b-6)c\zeta_\pm/3}e^{-\phi} 
\left ( \f{f_{,\zeta}}{3f} + \f{ac}{6} \right ) g_{\theta\theta}
\eea
Clearly, for a brane solution (energy = tension) we require $ac=0$. 
The Israel equations then give
\be
3\alpha[\cosh(c\beta\zeta) - \beta^{-1} \sinh(c\beta\zeta) ]
= 3\alpha \label{hetconstraint}
\ee
at either brane. (The sign of the energy term is taken care of by a flip
in the sign of the extrinsic curvature due to the normal pointing outwards
rather than inwards.) Obviously if $c=0$ this is trivially satisfied, but if
instead $a=0$, there is only one solution to (\ref{hetconstraint}), $\zeta=0$,
hence it is not possible to have the two brane heterotic set-up.

Therefore we conclude that for the two brane spacetime, we require the
bulk solution with $c=0$, and so the full spacetime is given by
(\ref{brnbulksoln}), the scalar field by (\ref{brnbulksclr}), and
the branes can be set at any fixed $y$-coordinate, which we will once more
take as $y = \pm y_0$ to compare with the background LOSW vacuum.
Restricting to the $+$ brane, the braneworld solution is
\bea
ds^2 &=& \left ( 1-\f{2E}{\rho} \right )^{\f{a}{E}} dt^2
- \left ( 1-\f{2E}{\rho} \right )^{-\f{a(1+b)}{E}}  [ d\rho^2 + \rho(\rho-2E)
d\Omega^2 ] \\
e^{2\phi} &=&  \left ( 1-\f{2E}{\rho} \right )^{\f{ab}{E}}
\label{brnblkhol}
\eea
Note however that the interbrane distance is not a constant:
\be
D = \int_{-y_0}^{y_0} dy |g_{yy}|^{1/2} = 2y_0 V_s(\rho)^{\f{ab}{2E}}
\ee
For $ab>0$, the interbrane distance decreases as $\rho$ decreases, 
eventually closing off the extra dimension at $\rho = 2E$.
For $ab<0$ however, the reverse is true, the branes move apart until 
at $\rho=2E$ the transverse separation is infinite.
\FIGURE{
\label{fig:contours}
\includegraphics[width=7cm]{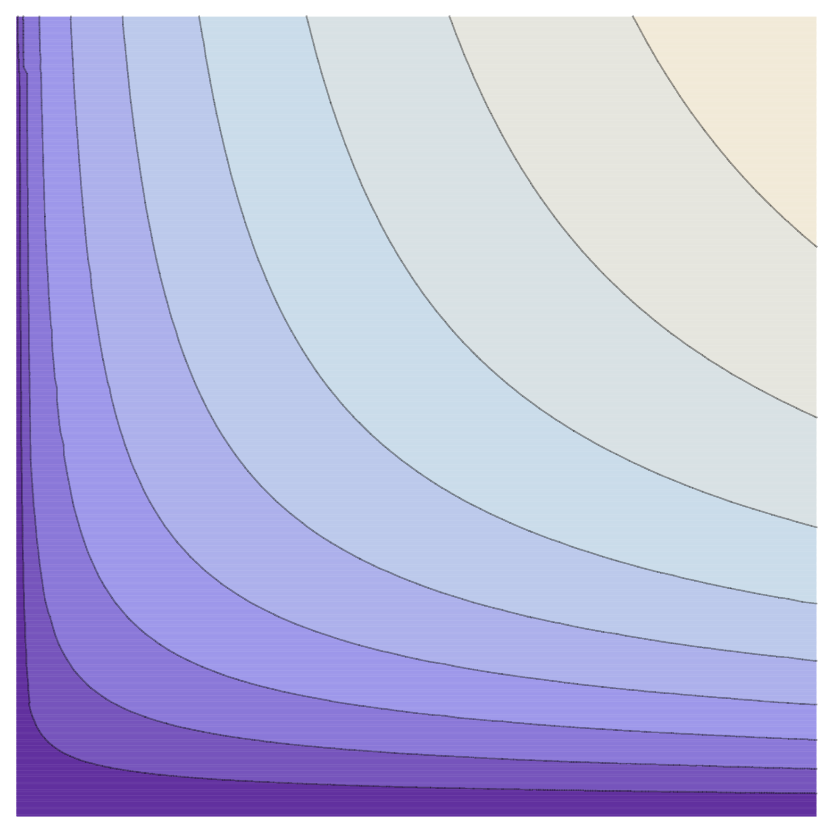}
\caption{A plot of the contours of $g_{tt}$ for the singular black 
tube solution ($\alpha = E = 1$). Note that far from the `horizon' 
the contours become
vertical, i.e.\ depend on the bulk transverse distance, but as $\rho$
decreases, the effect of the source begins to be felt.}}

Although either option yields a legitimate spherically symmetric
braneworld solution, a reasonable approach is to compare this exact
solution with the linearized result of the previous section:
\be
\delta \phi^+ = \f{2\alpha G_5 M}{r} \qquad
h^+_{tt} = -\f{10\alpha G_5 M}{r}
\ee
Expanding (\ref{brnblkhol}) at large $\rho$ yields:
\be
\phi \simeq - \f{ab}{\rho} \qquad
g_{tt} \simeq 1 - \f{2a}{\rho}
\ee
giving:
\be
a = 5\alpha G_5 M \qquad b = -2/5
\ee
Thus, matching to this linearized solution leads to a bulk in which the branes
become infinitely far apart as the null singularity is approached.

Thus, allowing for an axisymmetric bulk with two branes bounding it, and
assuming that the metric is separable, we have derived the general brane
radially symmetric solution which asymptotes the LOSW vacuum. Unfortunately
this solution is singular at $\rho=2E$, however, it does look
like the Schwarzschild solution at large $\rho$. The solution very much
resembles a string solution, however, the presence of the
Schwarzschild potential premultiplying the bulk $z$ coordinate causes
the interbrane distance to vary with $\rho$, and in fact the ``string''
becomes infinite as it becomes singular. 

\section{Discussion}

To sum up, we have explored the existence of braneworld black holes in the
heterotic braneworld scenario of Lukas, Ovrut, Stelle and Waldram.
We have shown how black string solutions are unstable, and that
the linearized solution has rather unusual asymptotics. Unfortunately
it was not possible to construct approximate brane stars, as the 
anisotropic nature of the LOSW vacuum means we have no spherically
symmetric bulk black hole solutions.  However, we were able to construct
an axisymmetric bulk solution which looks like Schwarzschild
at large distances, but which is singular as the Schwarzschild radius
is approached. 

Interestingly, the lack of a spherically
symmetric solution for any value of mass means that unlike the
RS and ADD models, we are unable to construct even a small black
hole perturbatively on the brane (such as the solutions considered in
\cite{SBH}) which seems somehow paradoxical as one might expect a small
black hole to be a small perturbation. However, this is really a signal
of the different bulk physics. In ADD and RS, the bulk is pure Einstein
gravity (with or without a cosmological constant) and at smaller scales
the brane becomes less and less relevant. In LOSW however, even a small
perturbation will interact with the scalar field, which is the breathing 
mode of the underlying Calabi-Yau manifold, and thus accesses
the higher dimensional physics this indicates.

The existence of these separable axisymmetric solutions is an
interesting consequence of the scalar field in the bulk, for the RS
model does not have an equivalent solution. The appearance of
the Schwarzschild potential to an irrational power is reminiscent of
the Poincare invariant $p$-brane solutions in pure gravity
\cite{RG}. In that case, a string solution in
higher dimensions was found, and while the solution was dependent
on only one variable (the radial distance) the effect of extra dimensions
was to introduce these irrational powers of the Schwarzschild potential.
Here we were looking for an {\it axi}-symmetric solution, with a warped
braneworld interpretation, yet, the effect of the extra dimension turns
out to be extremely similar.

The solution which corresponds to the braneworld linearized 
field at large $\rho$ has the branes diverging as we move in 
to smaller $\rho$. This is an extremely singular configuration 
with an infinite bulk null singularity. If however, we relax 
our requirements and do not demand agreement with the linearized solution,
then we can take $b>0$, in which case the bulk pinches off at $\rho=2E$, 
which is perhaps slightly preferable behaviour. Examining the linearized
scalar equation, (\ref{linscalar}), shows that for the branes to move
together, rather than apart, at the linearized level we require $T<0$,
in other words, $M < 3p$. This would mean matter with a stiff equation
of state, and unfortunately does not seem to match the separable solution.

Obviously the ansatz of separability was a choice, used in
order to get an exact analytic solution, and may be considered
to be too restrictive (although it is a common ansatz used in finding
supergravity solutions). Indeed, looking at the 
behaviour of the linearized solution across the bulk does not appear
to give the same dependence as the separable solution, although these
are in different gauges.  Clearly a numerical integration
would give a better indication of the true nature of the solution. However,
in spite of all the unattractive features, it is still interesting that
the heterotic braneworld does admit an analytic ``brane-vacuum'' spherically
symmetric solution, which
is the first example of an exact braneworld `black hole' solution
with a consistent bulk in 5 dimensions.

\section*{Acknowledgements}
We would like to thank Christos Charmousis for useful discussions. 
B.M. and A.L. acknowledge EPSRC and Durham University fellowships respectively. 


\end{document}